\documentclass[twocolumn,showpacs,preprintnumbers,amsmath,amssymb,superscriptaddress]{revtex4-1}

\usepackage[pdftex]{graphicx}
\usepackage{amsmath}

\begin{document}

\title{Linear magneto-electric phase in ultrathin MnPS$_{3}$ probed by optical second harmonic generation}

\author{Hao Chu}
\affiliation{Department of Physics, California Institute of Technology, Pasadena, CA 91125, USA}
\affiliation{Institute for Quantum Information and Matter, California Institute of Technology, Pasadena, CA 91125, USA}

\author{Chang Jae Roh}
\affiliation{Department of Physics and Photon Science, Gwangju Institute of Science and Technology, Gwangju 61005, Republic of Korea}

\author{Joshua O. Island}
\affiliation{Department of Physics, University of California, Santa Barbara, CA 93106, USA}

\author{Chen Li}
\affiliation{Department of Physics, California Institute of Technology, Pasadena, CA 91125, USA}
\affiliation{Institute for Quantum Information and Matter, California Institute of Technology, Pasadena, CA 91125, USA}

\author{Sungmin Lee}
\affiliation{Center for Correlated Electron Systems, Institute for Basic Science (IBS), Seoul 08826, Republic of Korea}
\affiliation{Department of Physics and Astronomy, Seoul National University (SNU), Seoul 08826, Republic of Korea}

\author{Jingjing Chen}
\affiliation{School of Physics, Nankai University, Tianjin 300071, China}

\author{Je-Geun Park}
\affiliation{Center for Correlated Electron Systems, Institute for Basic Science (IBS), Seoul 08826, Republic of Korea}
\affiliation{Department of Physics and Astronomy, Seoul National University (SNU), Seoul 08826, Republic of Korea}

\author{Andrea F. Young}
\affiliation{Department of Physics, University of California, Santa Barbara, CA 93106, USA}

\author{Jong Seok Lee}
\affiliation{Department of Physics and Photon Science, Gwangju Institute of Science and Technology, Gwangju 61005, Republic of Korea}

\author{David Hsieh}
\email[Author to whom correspondence should be addressed: ]{dhsieh@caltech.edu}
\affiliation{Department of Physics, California Institute of Technology, Pasadena, CA 91125, USA}
\affiliation{Institute for Quantum Information and Matter, California Institute of Technology, Pasadena, CA 91125, USA}

\date{\today}

\begin{abstract}
The transition metal thiophosphates $M$PS$_3$ ($M$ = Mn, Fe, Ni) are a class of van der Waals stacked insulating antiferromagnets that can be exfoliated down to the ultrathin limit. MnPS$_3$ is particularly interesting because its N$\acute{\textrm{e}}$el ordered state breaks both spatial-inversion and time-reversal symmetries, allowing for a linear magneto-electric phase that is rare among van der Waals materials. However, it is unknown whether this unique magnetic structure of bulk MnPS$_3$ remains stable in the ultrathin limit. Using optical second harmonic generation rotational anisotropy, we show that long-range linear magneto-electric type N$\acute{\textrm{e}}$el order in MnPS$_3$ persists down to at least 5.3 nm thickness. However an unusual mirror symmetry breaking develops in ultrathin samples on SiO$_2$ substrates that is absent in bulk materials, which is likely related to substrate induced strain.
\end{abstract}

\maketitle

Thin film materials that exhibit the magneto-electric (ME) effect - a coupling between magnetic (electric) polarization and external electric (magnetic) fields - have potentially broad applications in spintronics, sensing and energy harvesting technologies \cite{fiebig2005,spaldin}. Although ME effects in single-phase bulk crystals have been continuously pursued since their discovery in Cr$_2$O$_3$ in 1960 \cite{Dzyaloshinski,Astrov}, advances in thin film deposition techniques over the past two decades have opened new pathways to stabilize and to control high quality materials with large ME coupling strengths via epitaxial strain and heterostructure engineering, allowing the possibility of integration into functional nanoscale devices. At present, searching for both single-phase and composite thin film materials with stronger ME coupling, and developing methods to scale them down to the ultrathin few-unit-cell limit, remain active areas of research.

The recent discoveries of long-range magnetic ordering in exfoliated van der Waals (vdW) semiconductors \cite{huang2017,gong2017,burch,gibertini} potentially offer a new route to realizing ME materials in the ultrathin limit. The simplest type of ME effect, which involves a linear coupling between the external field and induced polarization, is allowed in materials that lack both spatial-inversion and time-reversal symmetries. As most of the naturally occurring vdW crystals are structurally centrosymmetric, a convenient strategy is to rely on the magnetic ordering itself to break inversion symmetry. This suggests that one should focus on antiferromagnetic (AF) rather than ferromagnetic (FM) materials because the latter generally do not break the inversion symmetry of the underlying lattice. It was recently reported that upon exfoliating CrI$_3$ down to a single bilayer, its magnetic order transforms from being FM to AF, breaking inversion symmetry and turning on a linear ME coupling in the process \cite{huang2017,huang2018,jiang2018,sun2019}. However, so far there are no reports of an ultrathin material that directly inherits linear ME properties from its bulk precursor.

The transition metal thiophosphates $M$PS$_3$ ($M$ = Mn, Fe, Ni) present an interesting family of AF vdW materials for such a study \cite{li2013a,sivadas2015,chittari2016,lee2016}. While the AF orders in FePS$_3$ and NiPS$_3$ preserve inversion symmetry \cite{wildes2015,lancon2016}, neutron diffraction studies have shown that the AF order in bulk MnPS$_3$ breaks inversion symmetry and allows a linear ME effect \cite{ressouche2010}. However, it is not clear if the linear ME-type AF order persists down to the ultrathin limit. Because such order does not exhibit any net magnetization, a magneto-optical Kerr rotation experiment is not applicable. Although Raman spectroscopy has detected phonon anomalies in ultrathin MnPS$_3$ that are potentially associated with AF ordering \cite{kim2019}, and spin transport measurements have shown evidence of persistent magnons in few layer MnPS$_3$ devices \cite{xing}, a technique that directly probes the AF structure in nanoscopic exfoliated samples is still urgently anticipated. Leveraging the sensitivity of optical second harmonic generation (SHG) to AF order \cite{fiebig_review}, we demonstrate here that SHG rotational anisotropy (RA) can directly couple to the AF order parameter in MnPS$_3$ nanoflakes, and use it to show that the linear ME-type AF order found in bulk MnPS$_3$ persists down to the ultrathin limit.

\begin{figure}[hbt!]
\centering
\includegraphics[width=0.425\textwidth]{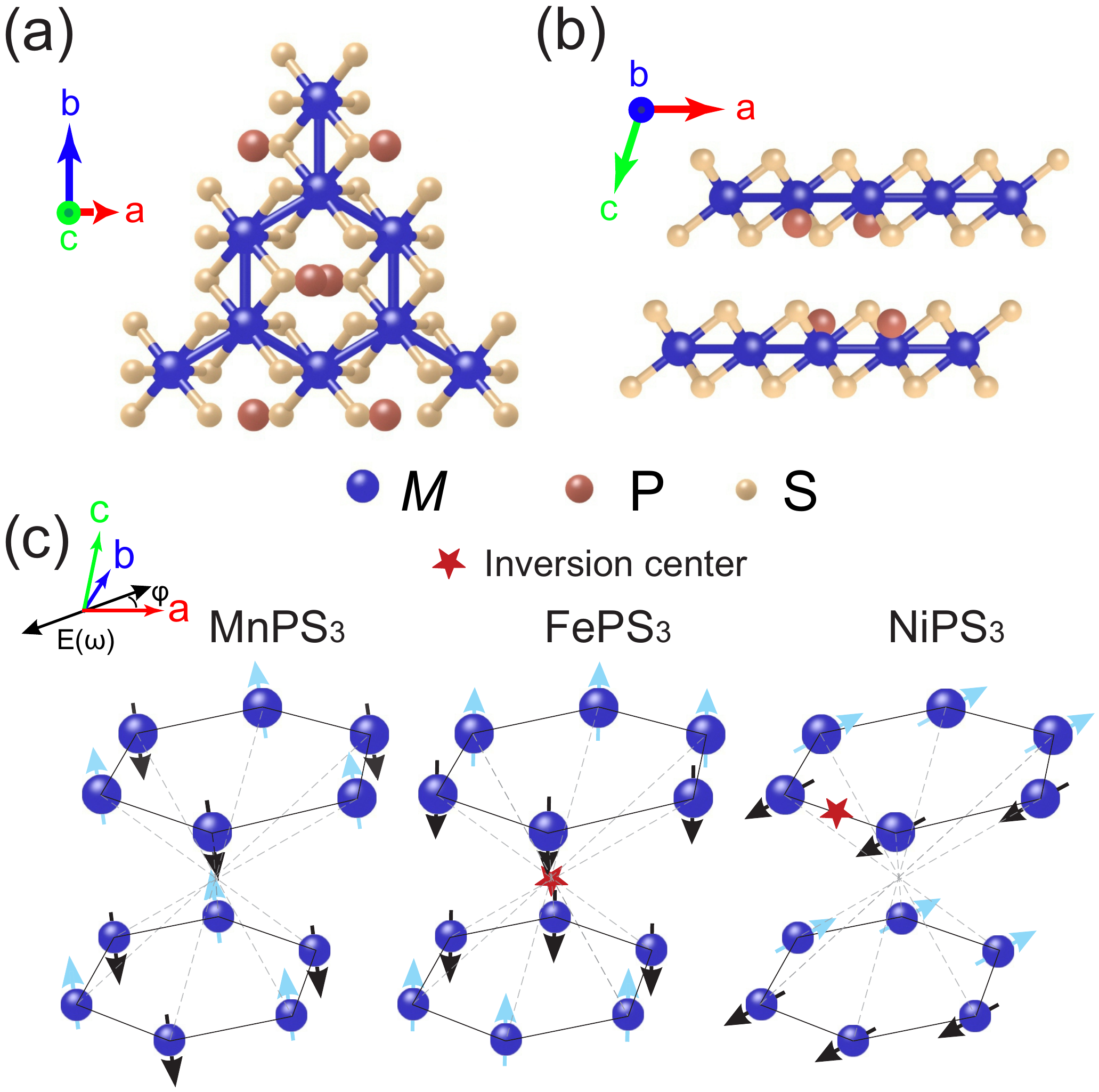}
\caption{Crystal and magnetic structure of $M$PS$_3$. $M$PS$_3$ lattice viewed along the (a) $c$- and (b) $b$-axis. Adjacent $ab$ planes are displaced by $a$/3 along the $\hat{a}$ direction. (c) AF structures of $M$PS$_3$. Arrows denote spin orientation. Star denotes an inversion center of the AF structure. The inset shows the in-plane orientation of the incident electric field.}\label{fig:logo}
\index{figures}
\end{figure}

Bulk MnPS$_3$ crystallizes in a monoclinic structure with centrosymmetric 2/$m$ point group symmetry \cite{ressouche2010}. It has a 2-fold rotation axis along the crystallographic $b$-axis and a mirror plane perpendicular to $\hat{b}$ [Fig. 1(a) \& (b)]. The Mn atoms are octahedrally coordinated by S atoms and form a honeycomb lattice in the $ab$ plane, but the in-plane 6-fold rotational symmetry of the honeycomb lattice is absent in the bulk crystal due to the displacement of adjacent layers along $\hat{a}$. Despite the similar lattice structures of MnPS$_{3}$, FePS$_{3}$ and NiPS$_{3}$ [Fig. 1(c)], MnPS$_3$ hosts an inversion broken N$\acute{\textrm{e}}$el-type AF order \cite{ressouche2010} whereas FePS$_3$ and NiPS$_3$ exhibit inversion symmetric zigzag-type AF order \cite{wildes2015,lancon2016}. The N$\acute{\textrm{e}}$el-type AF order preserves the size of the unit cell and exhibits no net moment, therefore it is challenging to detect via Raman spectroscopy and magneto-optical Kerr rotation respectively. However, because it breaks inversion symmetry, it should exhibit a finite second-order electric-dipole (ED) susceptibility that is responsible for SHG \cite{shen}. Therefore, we expect to see a finite SHG yield below the AF ordering temperature $T_{AF}$ from MnPS$_3$ and but not FePS$_{3}$ and NiPS$_{3}$.

\begin{figure}[hbt!]
\centering
\includegraphics[width=0.425\textwidth]{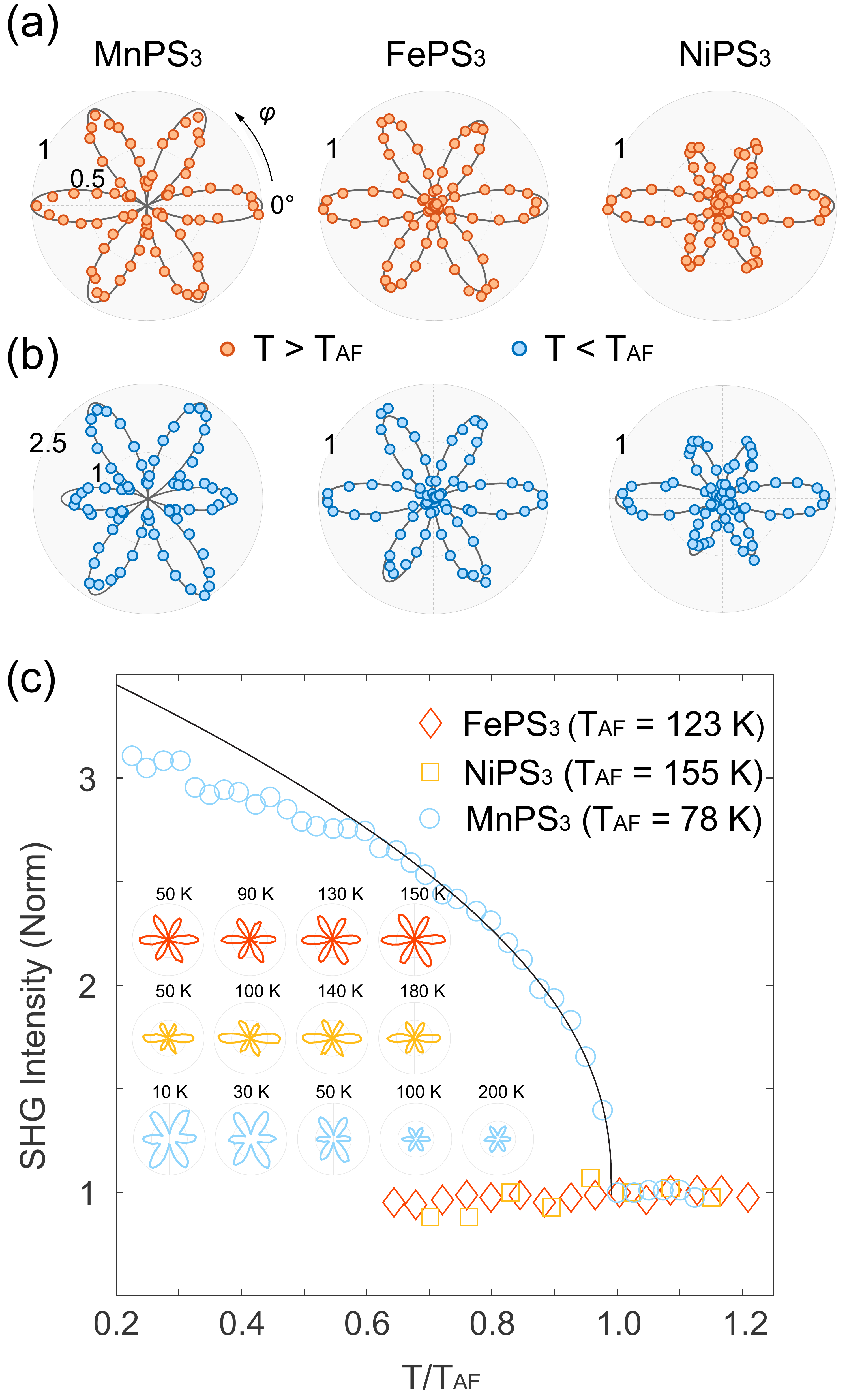}
\caption{SHG-RA patterns and long-range N$\acute{\textrm{e}}$el order in MnPS$_3$. (a) SHG-RA patterns of $M$PS$_3$ above and (b) below their respective AF ordering temperatures: $T_{AF}$ = 78 K (MnPS$_3$), 123 K (FePS$_3$) and 155 K (NiPS$_3$). Filled circles are experimental data and the solid lines are best fits to the phenomenological model described in the main text. For $T > T_{AF}$, all data were fit using only an EQ term. For $T < T_{AF}$, the FePS$_3$ and NiPS$_3$ data were fit using only an EQ term, whereas the MnPS$_3$ data were fit using a coherent sum of an EQ and ED term. (c) Temperature dependence of the SHG intensity along the $\phi$ = 60$^{\circ}$ direction. Solid line on the MnPS$_3$ data is a best fit to the power law function described in the main text, which accounts for a constant EQ term and a temperature dependent ED term.}\label{fig:logo}
\index{figures}
\end{figure}

The SHG-RA experiments were performed with a Ti:Sapph oscillator delivering laser pulses with a photon energy of $\hbar\omega$ = 1.5 eV, a pulse width of 80 fs, and a repetition rate of 80 MHz. The SHG photons produced at 3 eV are resonant with the band gap of MnPS$_3$ \cite{grasso}. A 5$\times$ (50$\times$) microscope objective was used to focus light onto the bulk (exfoliated) samples at normal incidence with a spot size of approximately 30 $\mu$m (2 $\mu$m), and the intensity of the reflected SHG beam was measured using a photomultiplier tube. The pulse energy of the incoming beam was kept below 50 pJ. The SHG-RA patterns were acquired by rotating the linear polarization of the incoming and outgoing beams (parameterized by the angle $\phi$), which were maintained parallel to each other, in the $ab$-plane [Fig. 1(c)]. Bulk $M$PS$_3$ single crystals were grown by a self-flux method described elsewhere \cite{lee2016a}.

\begin{figure}[hbt!]
\centering
\includegraphics[width=0.45\textwidth]{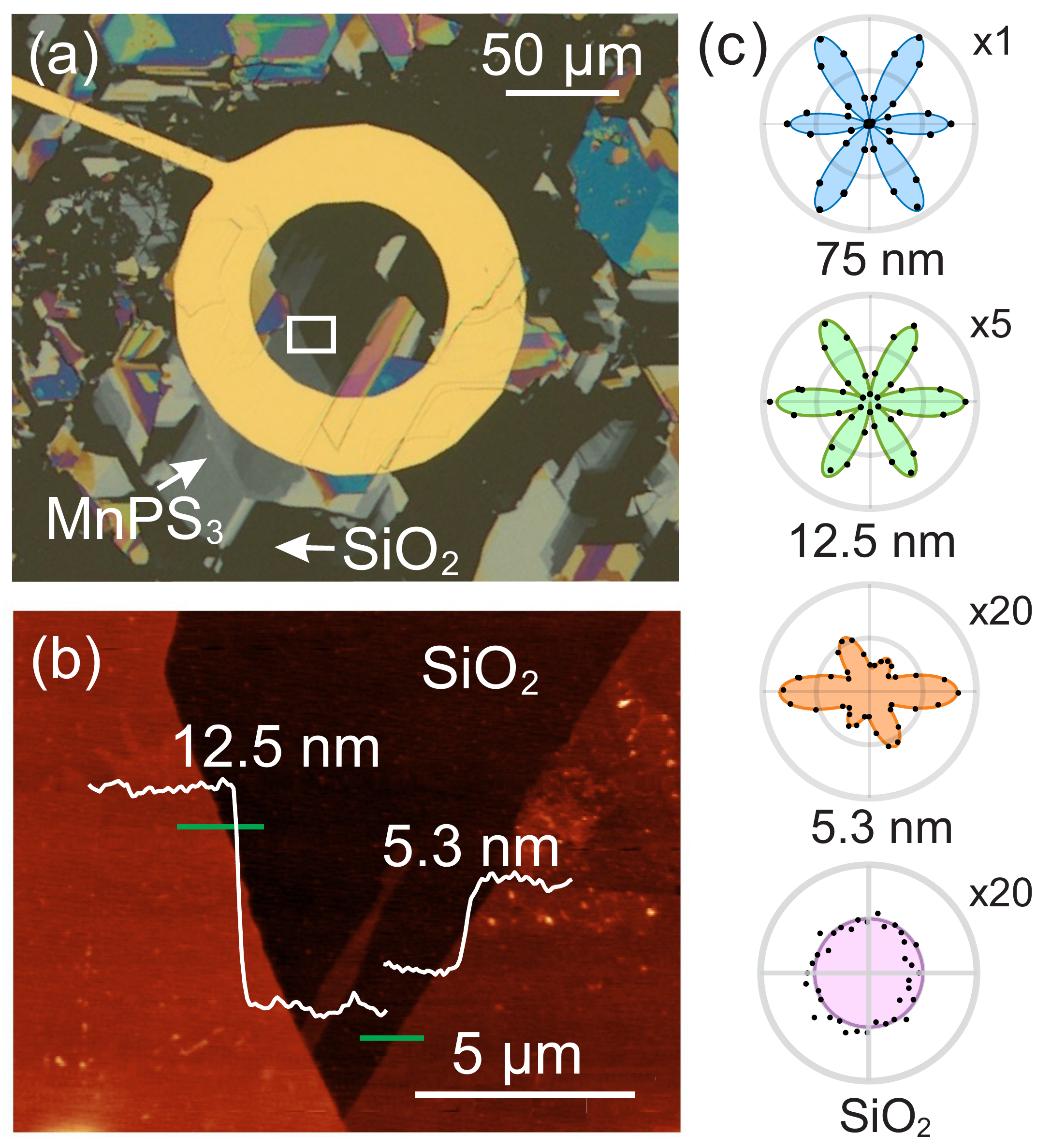}
\caption{MnPS$_3$ nanodevice for probing long-range N$\acute{\textrm{e}}$el order. (a) Optical image of exfoliated MnPS$_3$ flakes on SiO$_2$ with gold ring and electrode on top to improve cooling efficiency. The 5.3 nm and 12.5 nm flakes are found within the white box. (b) Atomic force microscopy scan of the area bounded by the white box in (a). Green lines indicate the positions of line scans, with corresponding magnified line profiles shown in white. (c) SHG-RA patterns from various regions of the device at 10 K.}\label{fig:logo}
\index{figures}
\end{figure}

Despite having a centrosymmetric crystallographic point group, we observe weak but finite SHG-RA signals from all three bulk crystals even above $T_{AF}$ [Fig. 2(a)]. This may arise from surface ED SHG or higher-rank bulk SHG processes such as electric-quadrupole (EQ) SHG \cite{shen}, both of which are generally allowed in centrosymmetric materials and were found to fit the data equally well [Fig. 2(a)] \cite{EPAPS}. For simplicity, we therefore only consider the bulk EQ term in our later fitting. The loss of 6-fold rotational symmetry that arises from the stacking offset between adjacent honeycomb layers is apparent in the data, although the degree of departure from 6-fold symmetry varies across samples as well as across spots within a single sample. We speculate that this may be due to spatial variations in the strength of inter-layer coupling and/or variations in the concentration of 120$^\circ$ twins or stacking faults \cite{murayama}.

Below $T_{AF}$ we observe no changes in the SHG intensity from both FePS$_{3}$ and NiPS$_{3}$, but an increase in the SHG intensity from MnPS$_{3}$ as anticipated. As shown in Fig. 2(b), the low temperature SHG-RA patterns from MnPS$_{3}$ can be well fit using the coherent sum of a non-magnetic EQ contribution and an AF order induced time-noninvariant ED contribution described by the equation \cite{EPAPS}:
\begin{equation}
P_{i}^{2\omega} = \chi^{EQ}_{ijkl} E_j^{\omega}\nabla_kE_l^{\omega}+\chi^{ED}_{ijk}(T) E_j^{\omega}E_k^{\omega}
\end{equation}
where $P^{2\omega}$ is the induced electric polarization at the SHG frequency, $E^{\omega}$ is the magnitude of the incident electric field, $\chi^{EQ}_{ijkl}$ is the temperature independent EQ susceptibility from a 2/$m$ crystallographic point group, and $\chi^{ED}_{ijk}(T)$ is a temperature dependent ED susceptibility from the 2$'/m$ magnetic point group describing the N$\acute{\textrm{e}}$el phase \cite{ressouche2010}. As shown in Fig. 2(c), the SHG intensity from MnPS$_{3}$ shows an order parameter-like increase below $T_{AF}$. Since $\chi^{ED}_{ijk}(T)$ is directly proportional to the inversion broken N$\acute{\textrm{e}}$el order parameter, we can extract the critical exponent of the order parameter ($\beta$) by fitting the temperature dependent SHG intensity to the phenomenological function $I^{2\omega}\propto[a + b(T_{AF}-T)^{\beta}]^2$, where $a$ is fixed by the intensity of the EQ contribution above $T_{AF}$ and both $b$ and $\beta$ are free parameters. Best fits to the region 60 K$ \leq T \leq T_{AF}$ yield $\beta=0.37(8)$ [Fig. 2(c)], which is close to the numerical calculation of $\sim$0.369 for the 3D Heisenberg model \cite{campostrini}.

\begin{figure*}[hbt!]
\centering
\includegraphics[scale=0.4,clip=true, viewport=0.0in 0in 17.1in 6.2in]{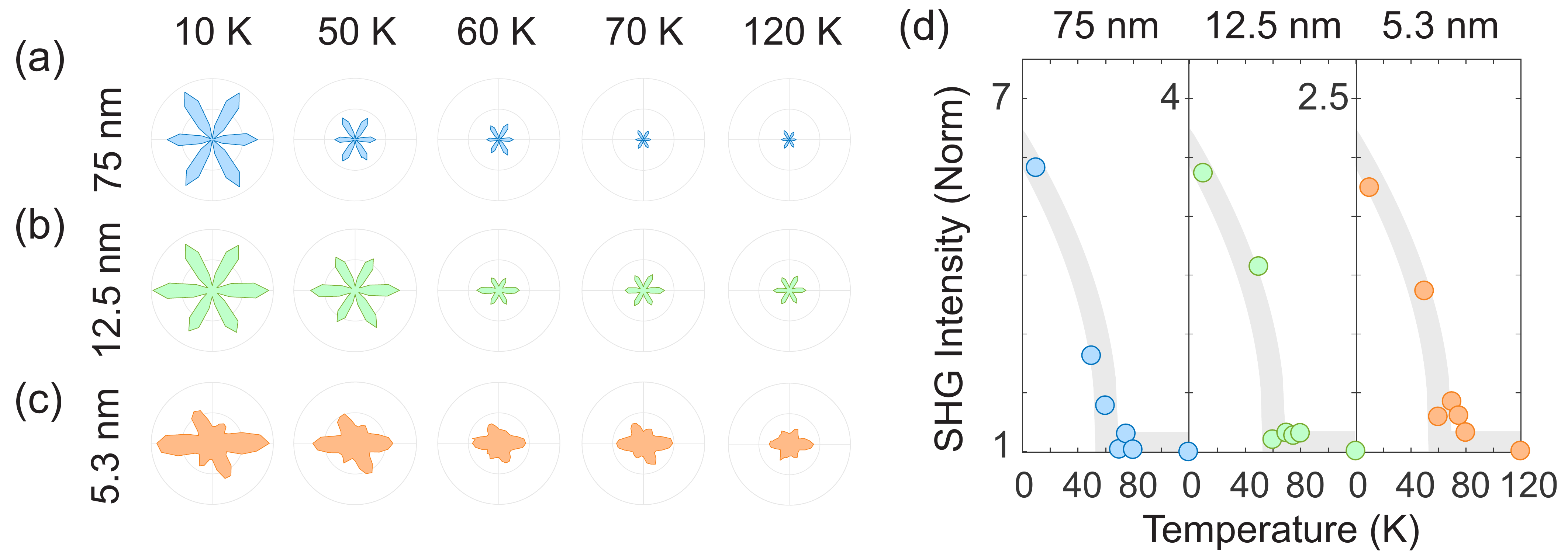}
\caption{Long-range N$\acute{\textrm{e}}$el order in few-layer MnPS$_3$. (a)-(c) Temperature dependent SHG-RA patterns from MnPS$_3$ flakes of different thicknesses. (d) Normalized temperature dependence of SHG Intensity along the $\phi$ = 0$^\circ$ direction.}\label{fig:logo}
\index{figures}
\end{figure*}

To investigate whether the long-range N$\acute{\textrm{e}}$el order in MnPS$_3$ survives in the ultrathin limit, we exfoliated bulk crystals onto an amorphous SiO$_2$ substrate in a nitrogen purged glove box. The choice of pure SiO$_2$ over SiO$_2$/Si as a substrate was made to reduce laser induced heating arising from optical absorption by Si at 800 nm and 400 nm. In contrast, SiO$_2$ is transparent to both 800 nm and 400 nm light. Due to the poor thermal conductivity of SiO$_2$ and the relatively high laser power needed for our SHG-RA measurements on MnPS$_3$ compared to other optical techniques for studying vdW magnets such as magneto-optical Kerr microscopy or Raman spectroscopy, we face more stringent sample cooling demands \cite{EPAPS}. To increase cooling efficiency, we deposited gold rings around the MnPS$_3$ flakes, which are thermally anchored to the cryostat sample holder by gold electrodes. Fig. 3(a) shows an optical image of a typical device. Using atomic force microscopy, we identified ultrathin MnPS$_3$ nanoflakes with 5.3 nm and 12.5 nm step sizes above the substrate on this device [Fig. 3(b)]. Based on previously published atomic force microscopy data on MnPS$_3$ \cite{lee2016a}, these correspond to 7 and 16 single layers of MnPS$_3$ respectively. Figure 3(c) shows typical SHG-RA patterns obtained from these flakes at a temperature of 10 K, compared with both thicker (75 nm) flakes and the bare substrate. We find that the overall SHG intensity approximately scales with the sample thickness, consistent with a bulk dominated SHG signal. The SiO$_2$ substrate contributes an isotropic background and is thus easily distinguished from the MnPS$_3$ signal.

As shown in Fig. 2(c), the ED SHG signal from MnPS$_3$ below $T_{AF}$ is of comparable magnitude to the high temperature EQ signal and is thus relatively weak overall. This is likely related to our incident 1.5 eV photon energy being well below the band gap ($\sim$ 3 eV) of MnPS$_3$. Consequently, when we attempted to protect the MnPS$_3$ flakes by encapsulation with a hexagonal boron nitride (hBN) thin flake, we found that the SHG signal was dominated by the hBN. Therefore we had to work with exposed MnPS$_3$ flakes, which are more prone to degradation. At cryogenic temperatures, we found that the SHG intensity from the few-layer regions starts to decrease over a time scale of several hours. This is likely due to surface adsorption of gas molecules and/or chemical reaction processes activated by laser exposure, as is observed in CrI$_3$ nanoflakes \cite{shcherbakov}. Therefore we were only able to acquire a limited number of SHG-RA scans at low temperatures before the onset of sample degradation. Nevertheless, our data clearly show an order parameter-like increase in the SHG intensity from the MnPS$_3$ nanoflakes below a temperature close to the bulk $T_{AF}$ value (Fig. 4), which again saturates at only several times the high temperature value. This indicates that the linear ME-type N$\acute{\textrm{e}}$el ordering observed in bulk crystals persists at least down to 7 layer thick samples. Measurements collected from 3 layer samples also show a markedly higher SHG intensity at 10 K compared to 100 K \cite{EPAPS}, but their faster degradation prevented a full temperature dependence measurement from being taken.

Given that the low temperature SHG signal from MnPS$_3$ involves the interference between a time-noninvariant ED response and a time-invariant EQ response, the existence of two different 180$^{\circ}$ AF domains related by time-reversal should produce different SHG intensities, analogous to what has previously been observed in Cr$_2$O$_3$ \cite{fiebig_cr2o3}. Assuming that the roughly 2-fold SHG intensity increase in our 5.3 nm flakes at low temperature [Fig. 4(d)] arises from a pure $+$ domain where the ED and EQ contributions interfere constructively, the ratio of the ED to EQ SHG electric fields should be roughly $\frac{1}{2}$. By extension, the SHG intensity from a $-$ domain is expected to be $\sim$ 25 \% of the high temperature value, or $(1+\frac{1}{2})^2/(1-\frac{1}{2})^2$ = 9 times lower in intensity compared to the $+$ domain. By raster scanning our beam of spot size $\sim$ 2 $\mu$m over the roughly 4 $\mu$m $\times$ 4 $\mu$m area of our 5.3 nm MnPS$_3$ flake at 10 K, we found the SHG intensity varies by only approximately $\pm$30 \% about the mean intensity. These small variations may be due to sample inhomogeneity and/or slight changes in alignment and are inconsistent with 180$^{\circ}$ domains. Therefore we believe our flake to be a single AF domain, which is comparable to the FM domain sizes observed in ultrathin vdW materials like CrI$_3$ \cite{huang2017} and Cr$_2$Ge$_2$Te$_6$ \cite{gong2017}.

We note that the low temperature SHG-RA patterns from the ultrathin 5.3 nm flakes exhibit an unusual symmetry. In particular, the $ac$ mirror plane (reflection about the horizontal line in the SHG-RA patterns) that is preserved by both the 2/$m$ and 2$'/m$ point groups is absent [Fig. 3(c)]. This mirror symmetry breaking only becomes apparent as the material thickness is reduced and as the temperature is lowered below $T_{AF}$ [Fig. 4(c)]. We believe that this is related to a substrate induced strain because for ultrathin MnPS$_3$ flakes that are exfoliated onto SiO$_2$/Si substrates, which are much smoother than pure SiO$_2$ substrates, there is no clear evidence of $ac$ mirror breaking in the low temperature SHG-RA patterns \cite{EPAPS}. Model Hamiltonian calculations \cite{cheng} show that a spiral spin texture that breaks $ac$ mirror symmetry is favored over the collinear N$\acute{\textrm{e}}$el order only if the second nearest-neighbor Dzyaloshinskii-Moriya interaction $D_2$ is comparable to the nearest-neighbor exchange $J_1$ in MnPS$_3$, which is around 1.5 meV according to inelastic neutron diffraction experiments \cite{wildes}. However spin Hall based measurements of $D_2$ in MnPS$_3$ put its value at merely 0.3 meV \cite{shiomi}. Since it is unlikely that $D_2$ is several times larger in ultrathin flakes compared to bulk crystals, especially given that Raman spectroscopy studies show no drastic changes in $T_{AF}$ or the phonon spectrum as a function of thickness \cite{kim2019}, we rule out a non-collinear spin texture as the cause for $ac$ mirror breaking. Instead, it is possible that the substrate induced strain tilts the easy-axis, causing the N$\acute{\textrm{e}}$el ordered moments to rigidly cant out of the $ac$ plane. Further structural and magnetic characterization of thin MnPS$_3$ flakes will be necessary to confirm this hypothesis.

In conclusion, we have demonstrated SHG-RA to be a direct and effective probe of inversion breaking AF order parameters in exfoliated vdW materials. A linear ME-type N$\acute{\textrm{e}}$el order that features in bulk crystals of MnPS$_3$ was found to survive down to the few layer limit. Future quantitative measurements of the ME coupling strength in ultrathin MnPS$_3$ samples will help to assess its potential for applications in nanoscale spintronics and optoelectronics devices.

\begin{acknowledgements}
Work at Caltech and UCSB was supported by ARO MURI Grant No. W911NF-16-1-0361. Work at GIST was supported by National Research Foundation of Korea (NRF) grants funded by the Korea government (MSIP) (No. 2018R1A2B2005331). Work at IBS CCES was supported by Institute for Basic Science (IBS) in Korea (IBS-R009-G1). D.H. and J.S.L also acknowledge support from a GIST-Caltech collaborative grant. J.O.I acknowledges the support of the Netherlands Organization for Scientific Research (NWO) through the Rubicon grant, project number 680-50-1525/2474.
\end{acknowledgements}

\end{document}